\documentstyle[preprint,aps,epsfig]{revtex}
\begin{document}
\draft
\title{Quasiparticle Energy Dispersion and Shadow Peaks
 in a Doped $\bbox{SO(5)}$ Symmetric Ladder}
\author{Seung-Pyo Hong and Sung-Ho Suck Salk}
\address{Department of Physics, Pohang University of Science and Technology,
 Pohang 790-784, Korea}
\maketitle
\begin{abstract}
Mean field and exact diagonalization studies on
the quasiparticle excitation of an $SO(5)$ symmetric two-leg ladder system
are reported. 
It is shown that 
the energy gap in the quasiparticle excitation is caused by
the formation of rung singlet states.
We find that shadow peaks can occur 
above the Fermi surface
with antiferromagnetic
electron correlations involving only rungs in the spin ladder. \\
\end{abstract}
\pacs{PACS numbers: 74.25.Jb, 74.25.-q, 71.10.-w, 71.27.+a}

\newpage
The cuprate materials of high critical superconducting temperatures exhibit 
antiferromagnetism near half-filling and
superconductivity away from half filling.
Zhang \cite{zhang97,zetal}
suggested that the two phenomena 
are manifestations of the same thing and 
can be explained in a unified framework 
based on $SO(5)$ symmetry.
Recently Scalapino et al.  \cite{szh97}
presented a simple $SO(5)$ model
on a two-leg ladder system
which can be served as a toy model to investigate
various physics involving the $SO(5)$ symmetry.
In this paper we discuss 
quasiparticle excitations for
the $SO(5)$ two-leg spin ladder
from both analytic and numerical results.
We derive a mean field quasiparticle energy dispersion relation.
For a through comparison
one-particle spectral function is calculated
by applying a Lanczos exact diagonalization method to
a ladder of $2\times6$ sites.
We find that the energy gap in the quasiparticle excitations
is largely contributed by the formation of rung singlet states.
We also find the appearance of shadow peaks
with Heisenberg interaction along rungs alone.

The $SO(5)$ symmetric two-leg ladder Hamiltonian is given by
\cite{szh97}
\begin{eqnarray}
 H &=&
 -t_\parallel \sum_{i\lambda\sigma}
 (c_{i\lambda\sigma}^\dag c_{i+1\lambda\sigma}^{} + \mbox{H.c.})
 -t_\perp \sum_{i\sigma}
 (c_{i1\sigma}^\dag c_{i2\sigma}^{} + \mbox{H.c.})
 \nonumber  \\
 && +U \sum_{i\lambda}
 \left(n_{i\lambda\uparrow} - \frac{1}{2} \right) 
 \left(n_{i\lambda\downarrow} - \frac{1}{2} \right) 
 +V \sum_{i}
 (n_{i1}-1) (n_{i2}-1) 
 \nonumber  \\
 && +J \sum_{i}
 {\bf S}_{i1} \cdot {\bf S}_{i2}
 -\mu \sum_{i\lambda\sigma}
  n_{i\lambda\sigma}
 \label{hamiltonian}
\end{eqnarray}
Here $c_{i\lambda\sigma}^\dag$ creates an electron with spin $\sigma$
on the $i$-th rung of the $\lambda$-th leg with $i=1,\ldots,L$ and
$\lambda=1,2$.
$n$ is the number operator,
$n_{i\lambda\sigma} = c_{i\lambda\sigma}^\dag c_{i\lambda\sigma}^{}$,
$n_{i\lambda} = \sum_\sigma n_{i\lambda\sigma}$,
and ${\bf S}$ is the spin operator,
${\bf S}_{i\lambda} = \frac{1}{2} c_{i\lambda\alpha}^\dag
 \mbox{\boldmath $\sigma$}_{\alpha\beta} c_{i\lambda\beta}^{}$.
$t_\parallel$ is the hopping integral in the leg direction;
$t_\perp$, the hopping integral in the rung direction;
$U$, the on-site Coulomb interaction;
$V$, the near-neighbor Coulomb interaction on a rung;
$J$, the Heisenberg exchange interaction on a rung;
and $\mu$, the chemical potential.
The Heisenberg interaction of the $SO(5)$ ladder Hamiltonian
allows both singly and doubly occupied sites,
while that of the $t$-$J$ ladder Hamiltonian
allows only singly occupied sites.
The constraint for the interaction strengths,
$J=4(U+V)$ in Eq. (\ref{hamiltonian}) above
is required for the $SO(5)$ symmetry.
$U$ and $V$ can be repulsive or attractive potentials.
In the present study
we will consider only the case of repulsive potentials, $U\ge 0$ and $V\ge 0$,
which is in the region of the $E_0$ spin gap phase 
(represented by the tensor product of rung singlet states 
in the strong coupling limit)
defined by Scalapino et al. \cite{szh97}

The weak coupling (that is, $U,V \ll t_\parallel \simeq t_\perp$)
may allow linearization of the Coulomb repulsion terms
\begin{eqnarray*}
 \lefteqn{
 \left( n_{i\lambda\uparrow} - \frac{1}{2} \right)
 \left( n_{i\lambda\downarrow} - \frac{1}{2} \right)
 }
 \nonumber \\
 &=&
 \left\langle n_{i\lambda\uparrow} - \frac{1}{2} \right\rangle
 \left( n_{i\lambda\downarrow} - \frac{1}{2} \right)
 +
 \left\langle n_{i\lambda\downarrow} - \frac{1}{2} \right\rangle
 \left( n_{i\lambda\uparrow} - \frac{1}{2} \right)
 - 
 \left\langle n_{i\lambda\uparrow} - \frac{1}{2} \right\rangle
 \left\langle n_{i\lambda\downarrow} - \frac{1}{2} \right\rangle
\end{eqnarray*}
and
\begin{eqnarray*}
 \lefteqn{
 (n_{i1}-1) (n_{i2}-1)
 }
 \nonumber \\
 &=&
   \langle n_{i1}-1 \rangle (n_{i2}-1)
 + \langle n_{i2}-1 \rangle (n_{i1}-1)
 - \langle n_{i1}-1 \rangle \langle n_{i2}-1 \rangle 
\end{eqnarray*}
Taking into account
$\langle n_{i\lambda\uparrow} \rangle
 = \langle n_{i\lambda\downarrow} \rangle
 = \frac{1-\delta}{2}$
with $\delta$, the doping rate,
which neglects hole density fluctuations,
we write the Coulomb repulsions 
$( n_{i\lambda\uparrow} - \frac{1}{2} )
 ( n_{i\lambda\downarrow} - \frac{1}{2} ) 
  = -\frac{\delta}{2} n_{i\lambda} +\frac{\delta}{2} - \frac{\delta^2}{4}$
and
$(n_{i1}-1) (n_{i2}-1) = -\delta(n_{i1}+n_{i2}) + 2\delta - \delta^2$,
and add them to the chemical potential term
in Eq. (\ref{hamiltonian}).
The Heisenberg interaction can be written 
\cite{ul92} as
\begin{eqnarray*}
 {\bf S}_{i1} \cdot {\bf S}_{i2}
 &=&
 -\frac{3}{8} [\chi_{i12}^\ast 
 ( c_{i1\uparrow}^\dag c_{i2\uparrow}^{}
 + c_{i1\downarrow}^\dag c_{i2\downarrow}^{} )
 + \mbox{H.c.} ]
 +\frac{3}{8} n_{i1}
 \nonumber \\
 &&
  -\frac{3}{8} [\Delta_{i12}^\ast 
 ( c_{i1\uparrow}^{} c_{i2\downarrow}^{}
 - c_{i1\downarrow}^{} c_{i2\uparrow}^{} )
 + \mbox{H.c.} ]
\end{eqnarray*}
where
the hopping order parameter is
$\chi_{i12}^{} = \langle c_{i1\uparrow}^\dag c_{i2\uparrow}^{}
 + c_{i1\downarrow}^\dag c_{i2\downarrow}^{} \rangle$
and
the singlet pair order parameter,
$\Delta_{i12}^{} = \langle c_{i1\uparrow}^{} c_{i2\downarrow}^{}
 - c_{i1\downarrow}^{} c_{i2\uparrow}^{} \rangle$.
We take $\chi_{i12}^{} = \chi$ and $\Delta_{i12}^{} = \Delta$
by neglecting spatial fluctuations
of both the amplitude and the phase.
The mean field Hamiltonian is then in momentum space,
\begin{eqnarray}
 H
 &=&
 \sum_{k\lambda\sigma}
 (-2 t_\parallel\cos k_x - \mu)
 c_{k\lambda\sigma}^\dag c_{k\lambda\sigma}^{}
 - t_\perp \sum_{k\sigma}
 (c_{k1\sigma}^\dag c_{k2\sigma}^{} + \mbox{H.c.})
 \nonumber \\
 &&
 -\frac{3J}{8} \sum_{k}
 [ \chi^\ast (c_{k1\uparrow}^\dag c_{k2\uparrow}^{}
  + c_{k1\downarrow}^\dag c_{k2\downarrow}^{})
  + \Delta^\ast ( c_{k1\uparrow}^{} c_{-k2\downarrow}^{}
  - c_{k1\downarrow}^{} c_{-k2\uparrow}^{} )
  + \mbox{H.c.} ]
 \label{meanham}
\end{eqnarray}
The quasiparticle energy dispersion is readily obtained 
from Eq. (\ref{meanham}) above,
\begin{equation}
 E_k = \pm \sqrt{\left[ (-2t_\parallel\cos k_x - \mu) \pm
 \left( t_\perp+\frac{3J\chi}{8}\right) \right]^2 
 + \left(\frac{3J\Delta}{8}\right)^2}
 \label{Ek}
\end{equation}

\begin{figure}
 \vspace*{40mm}
 \centering
 \epsfig{file=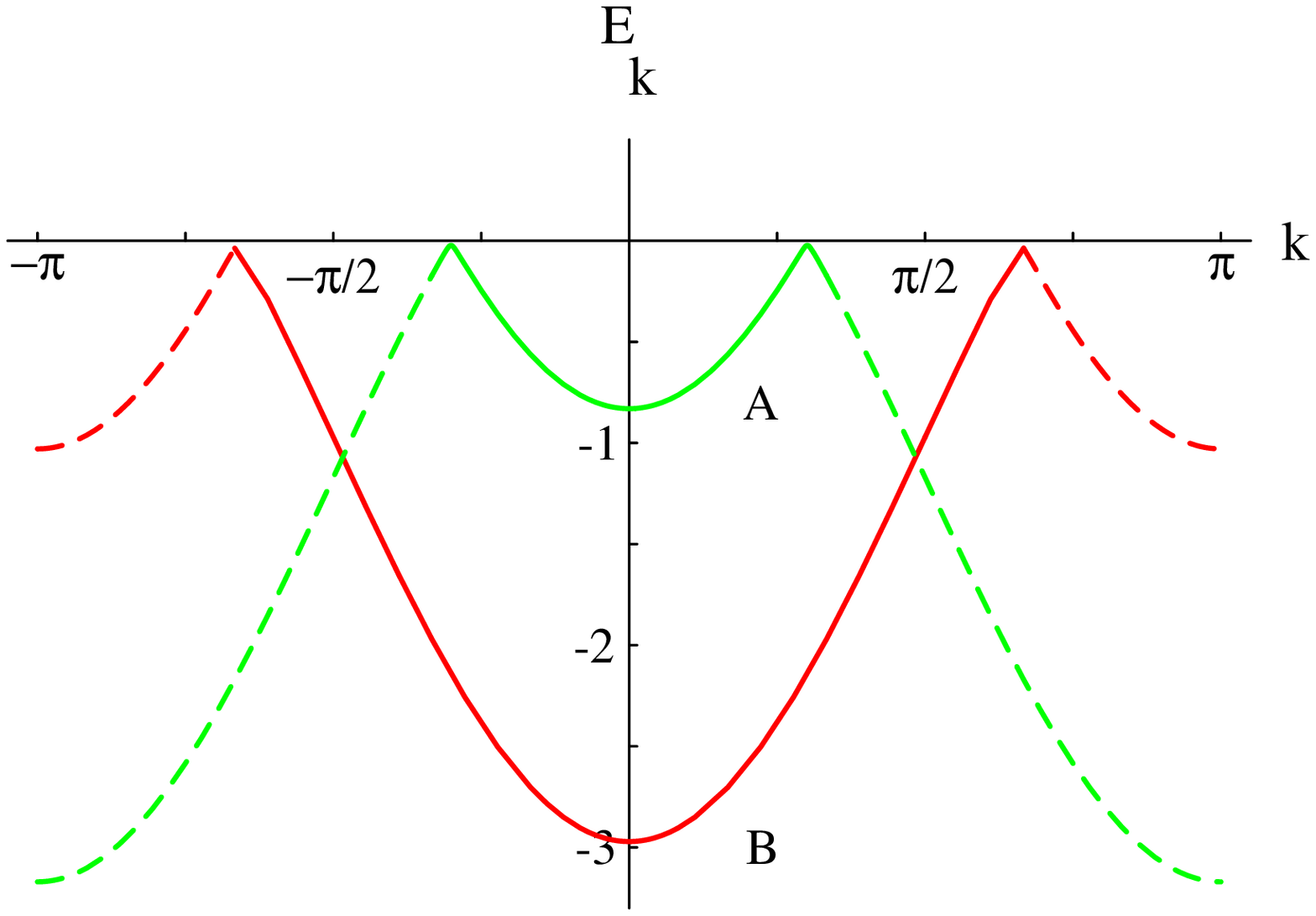, width=13cm}
 \caption{}
 \label{dispersion}
\end{figure}

The mean field order parameters $\chi$, $\Delta$, 
and the chemical potential $\mu$ are 
obtained from self-consistent equations.
For the time being we take the values of
$\chi$, $\Delta$, and $\mu$ obtained from our
Lanczos calculations on the $2\times6$ ladder with two doped holes.
The quasiparticle energy dispersion is shown in Fig. \ref{dispersion};
the bonding band (denoted as B)
has its minimum energy at $k_x=0$
with the Fermi surface at $k_F^B \simeq \frac{2\pi}{3}$,
while the antibonding band (denoted as A)
has its minimum energy at $k_x=0$
with the Fermi surface at $k_F^A \simeq \frac{\pi}{3}$.
Splitting between the two bands,
that is, the removal of degeneracy is caused by 
hopping (overlap) integral $t_\perp$
in the rung direction.
The bonding band is pushed down with more occupied electrons,
while the antibonding band is pushed up with less occupied electrons.
We note that with $k_F^B > k_F^A$
the Luttinger sum rule is satisfied, that is,
$k_F^B + k_F^A = (1-\delta)\pi$.
The dashed parts of the dispersion curves 
in Fig. \ref{dispersion} represent the shadow bands
in our two-leg ladder system
which is similar to the ones
discussed in two-dimensional planar systems
\cite{ks90,aosssmk94,hmd95}.
As shown in Eq. (\ref{Ek})
the shadow band in Fig. \ref{dispersion}
is largely contributed by the Heisenberg interaction in the rung direction,
while the shadow band in the two-dimensional planar systems
\cite{ks90,aosssmk94,hmd95}
is caused by antiferromagnetic correlations
whose strengths are equal in both the horizontal and vertical directions.

\begin{figure}
 \centering
 \epsfig{file=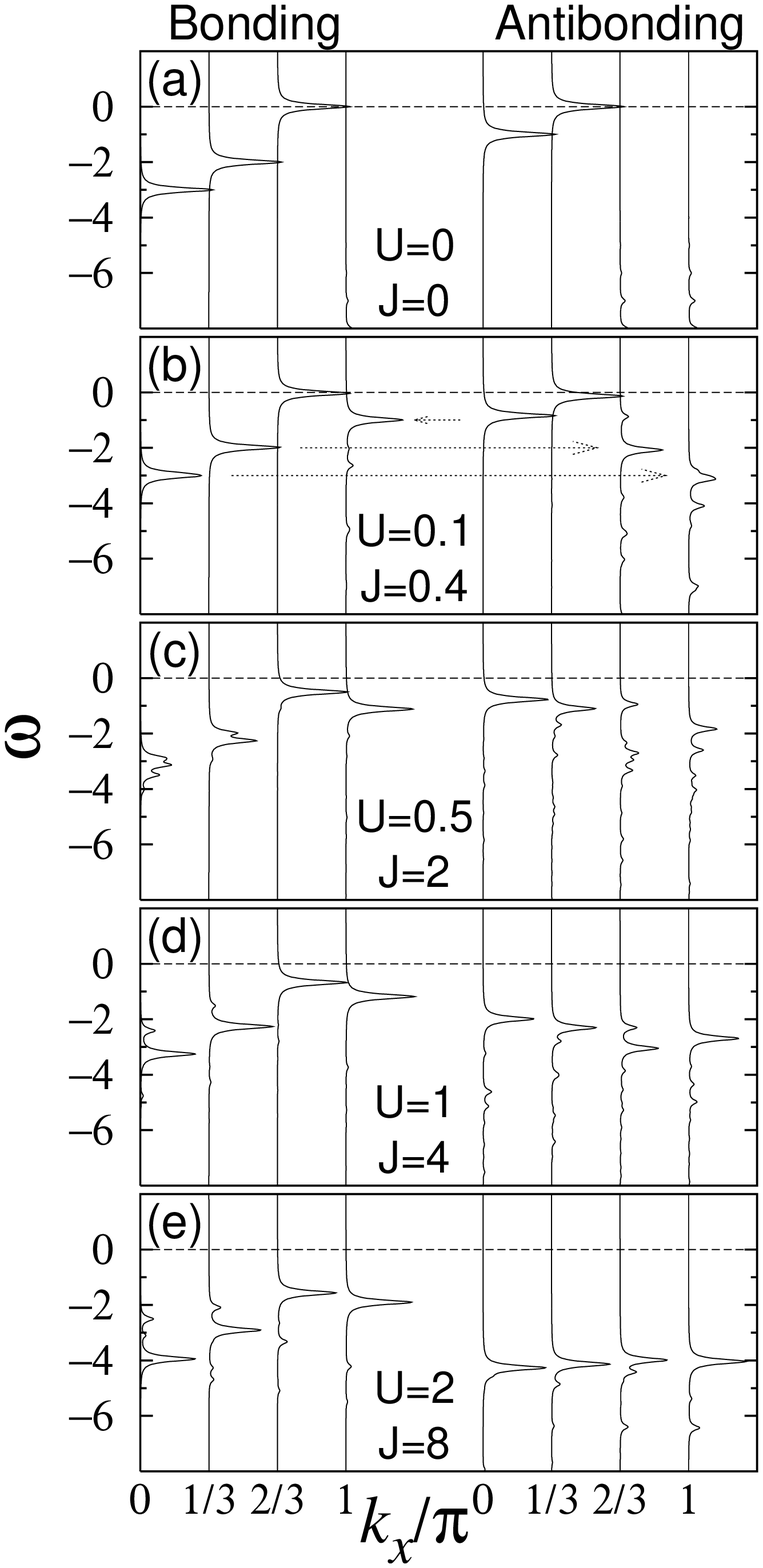, width=11cm}
 \caption{}
 \label{spectrumuj}
\end{figure}

To examine the single particle excitations,
we write in the case of two-hole doped system of $2\times L$ sites
\cite{ttr94},
\addtocounter{equation}{1}
$$
 A_e({\bf k}, \omega)
 =
 \sum_\alpha \left| \langle \Psi_\alpha^{2L-1}|c_{{\bf k}\sigma}^\dag
 | \Psi_0^{2L-2}\rangle \right|^2
 \delta(~~\omega - E_\alpha^{2L-1} + E_0^{2L-2} + \mu) 
 \eqno{(\arabic{equation}.a)}
$$
for the particle spectral function $(\omega>0)$ and
$$
 A_h({\bf k}, \omega)
 =
 \sum_\alpha \left| \langle \Psi_\alpha^{2L-3}|c_{{\bf k}\sigma}^{}
 | \Psi_0^{2L-2}\rangle \right|^2
 \delta(-\omega - E_\alpha^{2L-3} + E_0^{2L-2} - \mu) 
 \eqno{(\arabic{equation}.b)}
$$
for the hole spectral function $(\omega<0)$.
Here $|\Psi_\alpha^N\rangle$ is the $\alpha$-th eigenstate
in the subspace of $N$ electrons with the eigenenergy $E_\alpha^N$.
The chemical potential is defined as
$\mu=\frac{1}{2}(E_0^{2L-1} - E_0^{2L-3})$.
We calculate the spectral function 
by using both the Lanczos method
and the continued fraction approach \cite{dagotto94}.
The results are shown in Fig. \ref{spectrumuj}
for $L=6$, $t_\parallel=t_\perp=1$ and $V=0$,
as a function of $U$ and $J$.
The coupling strengths, $U$ and $J$ are related to each other and
$J$ increases faster than $U$
due to
the $SO(5)$ constraint, $J=4(U+V)=4U$.

We now investigate the spectral functions 
by varying the values of $U$ (and consequently $J$).
First the computed free particle spectral functions 
for the case of $U=J=0$ are shown in Fig. \ref{spectrumuj}(a).
The Fermi momentum of the bonding $(k_y=0)$ band is
$k_F^B = \frac{2\pi}{3}$,
and that of the antibonding $(k_y=\pi)$ band is 
$k_F^A = \frac{\pi}{3}$.
Below the Fermi surface,
we observe
sharp and well-defined quasiparticle peaks
for both the bonding and the antibonding bands.
All the single particle states below the Fermi surface are occupied,
and there exist no spectral peaks above the Fermi surface.
The computed spectral functions 
shown in Fig. \ref{spectrumuj}(a)
are seen to agree well with the free particle energy dispersion,
$\varepsilon_k^{} = -2t_\parallel \cos k_x \pm t_\perp$ in Eq. (\ref{Ek}).

The spectral functions for the case of small coupling strengths,
$U=0.1$ and $J=0.4$
are displayed in Fig. \ref{spectrumuj}(b).
The predicted Fermi momenta are found at
$k_F^B \simeq \frac{2\pi}{3}$
and $k_F^A \simeq \frac{\pi}{3}$
in agreement with the mean field result 
given by Eq. (\ref{Ek}).
The calculated positions of the quasiparticle peaks 
below the Fermi surface
are also in good agreement with the analytic mean field result
of Eq. (\ref{Ek}).
Although the interaction strengths are small,
interestingly enough there appear spectral peaks
above the Fermi momentum (surface).
Such peaks are the shadow peaks 
which were also observed by others \cite{hd96}.
The peak in the bonding band with momentum $(\pi,0)$ above the Fermi surface
is a shadow of the peak belonging to 
the antibonding band with momentum $(0,\pi)$
which is below the Fermi surface.
The peak with momentum $(\frac{2\pi}{3}, \pi)$
is a shadow of the peak with momentum $(-\frac{\pi}{3},0)$
(degenerate at $(\frac{\pi}{3},0)$),
and the peak with momentum $(\pi,\pi)$ is
a shadow of the peak with momentum $(0,0)$.
Haas and Dagotto \cite{hd96} reported 
the presence of shadow peaks in $t$-$J$ ladder systems
and concluded that 
short-range antiferromagnetic correlations
are responsible for the shadow peaks
in the spin ladder system.

\begin{figure}
 \centering
 \epsfig{file=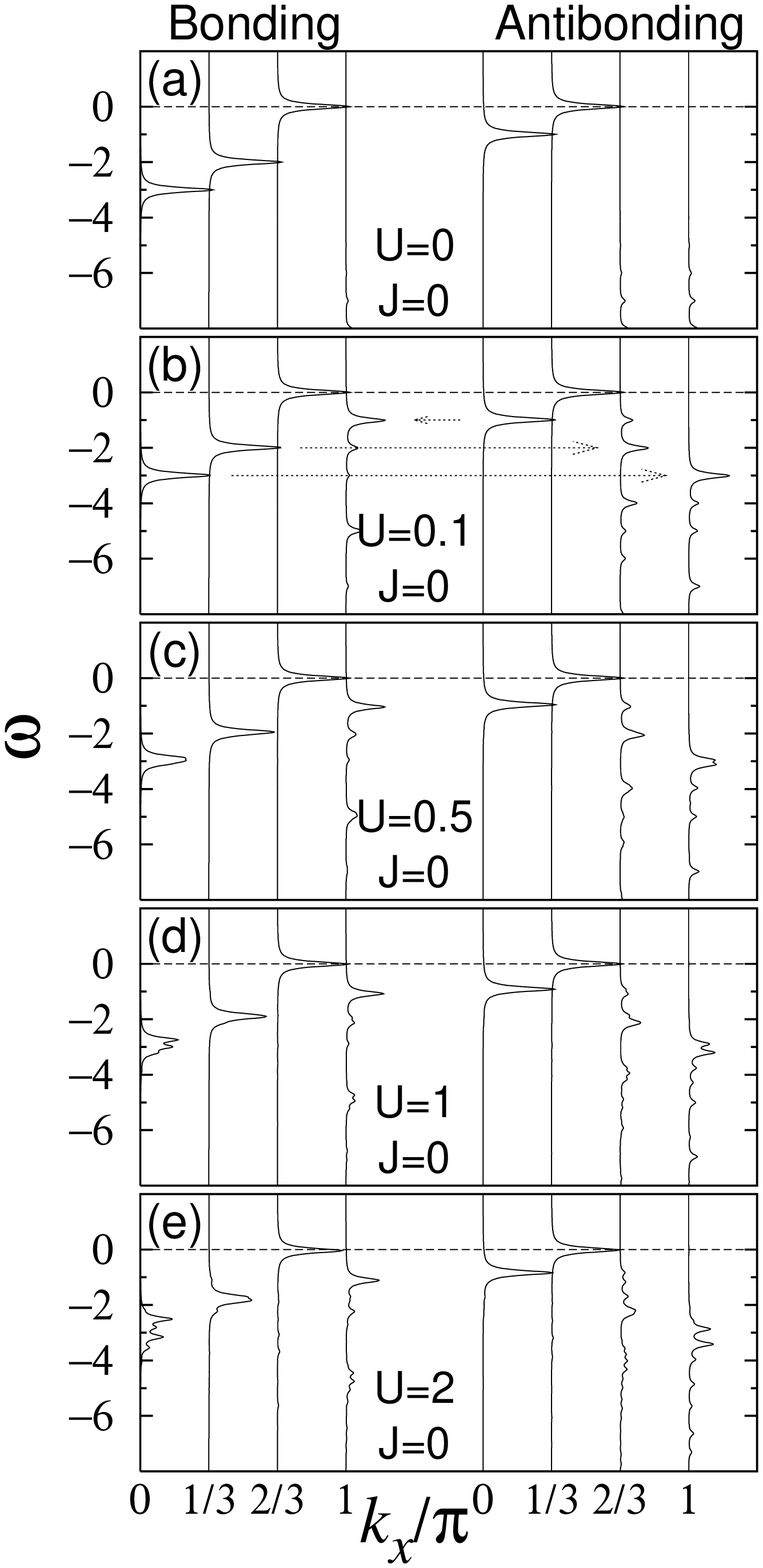, width=11cm}
 \caption{}
 \label{spectrumu}
\end{figure}

In order to investigate
which of the two parameters
$U$ and $J$ is more responsible for yielding the shadow peaks,
we first set $J=0$
and calculate spectral functions for several cases of $U$
as shown in Fig. \ref{spectrumu}.
We note that
the stronger the on-site Coulomb repulsion,
the broader the bonding orbital becomes,
which is more evident for the orbitals,
particularly at lower values of $k_x$
(substantially below the Fermi surface),
for example, at $k=(0,0)$ and $(\frac{\pi}{3},0)$
as shown in Fig. \ref{spectrumu}.
The overall structure of the bonding and the antibonding bands
does not change much with increasing $U$.
We find that the shadow peaks can be generated 
even with a small Coulomb repulsion $U=0.1$
as shown in Fig. \ref{spectrumu}(b).
While Haas and Dagotto's results \cite{hd96} 
are based on the strong coupling limit $U \gg t$
due to the use of the $t$-$J$ Hamiltonian,
we discover the shadow peaks in both the weak and the strong coupling limits.
Second order hopping processes generates an effective interaction
of strength $\sim t^2/U$
between the electrons on nearest-neighbor sites.
Since the hopping integral along the rung ($t_\perp$)
and that along the chain  ($t_\parallel$)
are the same,
it is hard to determine from their study which direction
of electron correlations is more important
for the shadow peaks.

\begin{figure}
 \centering
 \epsfig{file=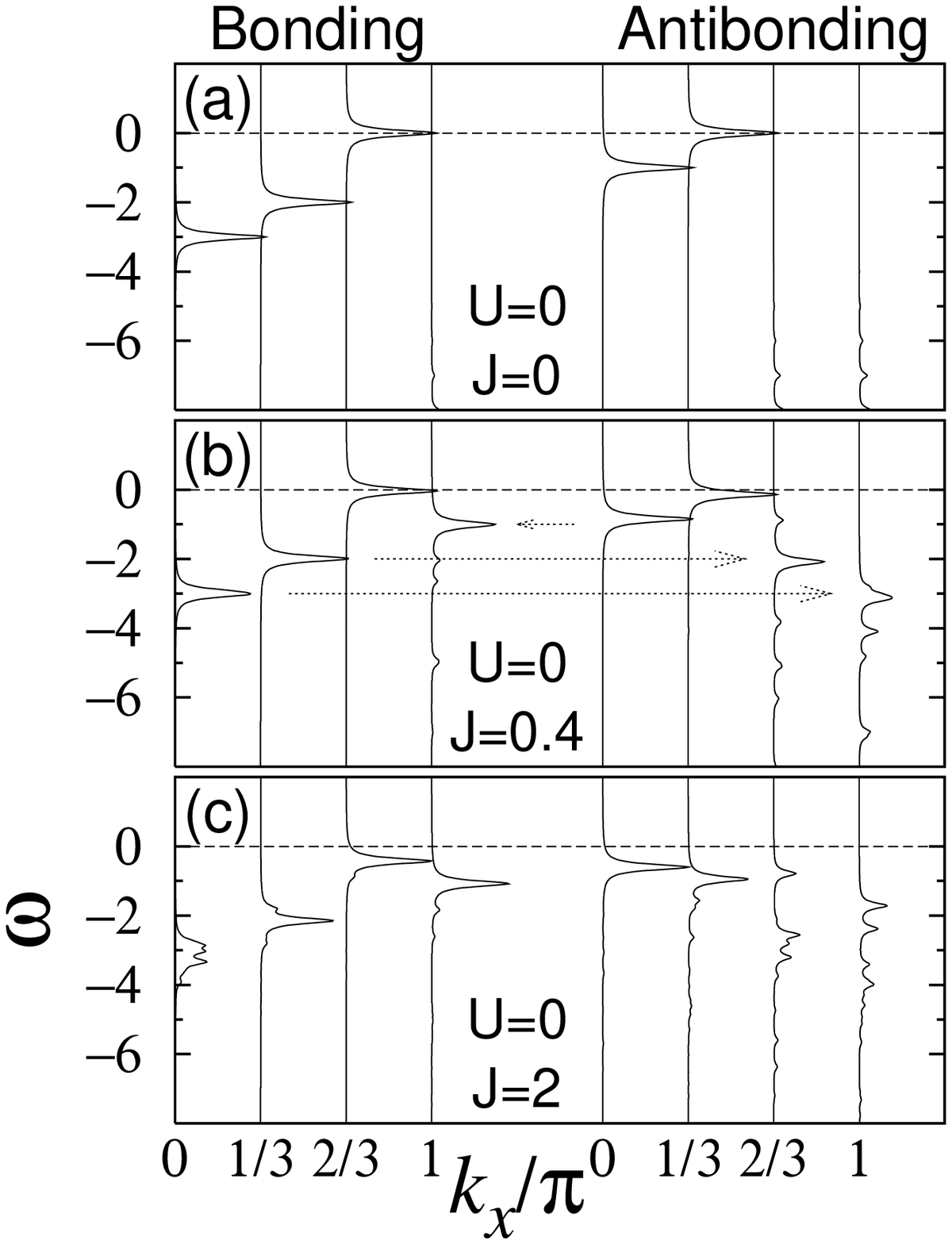, width=11cm}
 \caption{}
 \label{spectrumj}
\end{figure}

In order to thoroughly verify the validity of shadow peaks caused
essentially by the antiferromagnetic correlation 
between the two electrons in the rung,
namely, by the spin singlet state along the rung direction,
we computed the spectral functions 
for the case of $U=V=0$, $J_\perp \equiv J \ne 0$ and $J_\parallel =0$
as shown in Fig. \ref{spectrumj}.
We find that the shadow peaks appear above the Fermi surface
with the Heisenberg interaction along rungs alone
as displayed in Fig. \ref{spectrumj}(b).
Thus the singlet bonding on the rungs is responsible for
the presence of the shadow peaks.

Now we reexamine the $SO(5)$ symmetric cases of $J=4U$.
The spectral functions for different values of $U$ and $J$
are shown in Figs. \ref{spectrumuj}(c),(d),(e).
As $J$ increases,
the energies of the bonding and the antibonding orbitals decrease
with increasing energy gap.
It is to be reminded that
due to the $SO(5)$ constraint to define the relation between $J$ and $U$,
$J$ increases more rapidly (4 times faster) than $U$.
Thus it is of great interest 
to see how the spectral positions and energy gaps vary
as a result of larger contribution by $J$.
As can be seen from the comparison of 
Figs. \ref{spectrumuj} and \ref{spectrumu},
we note that the energy gap is largely
contributed by $J$,
but not substantially so by $U$.
The dominance of spin singlet states on the rungs
for the determination of the energy gap
is manifest from this study.
Indeed this observation is in good accordance with
the mean field prediction;
the energy gap is caused by the Heisenberg interaction $J$
as shown in Eq. (\ref{Ek}).

The dispersion of the bonding band does not greatly change
with increasing values of $J$,
in contrast to that of the antibonding band.
As $J$ increases,
the antibonding band is lowered more rapidly than the bonding band
and finally becomes dispersionless (no dependence on $k_x$)
as shown in Fig. \ref{spectrumuj}(e).
The Heisenberg interaction $J$,
which is taken to be anisotropic
due to $J_\perp \ne 0$ and $J_\parallel=0$,
obviously favors the formation of the local rung singlet.
The rung singlets are clearly local
because
there exists no phase coherence 
between the singlets on the adjacent rungs,
which, in turn, causes the appearance of 
the dispersionless antibonding band.

We have investigated quasiparticle excitations
in the $SO(5)$ symmetric two-leg ladder system
recently proposed by Scalapino et al. \cite{szh97}
The quasiparticle energy dispersion is discussed
in both the mean field approach
and the exact diagonalization study.
The dispersion consists of two branches,
the bonding band with Fermi momentum $k_F^B \simeq \frac{2\pi}{3}$
and the antibonding band with Fermi momentum $k_F^A \simeq \frac{\pi}{3}$.
The hole spectral functions calculated 
from the Lanczos exact diagonalization method
were found to agree well with the mean field results.
The formation of rung singlet states
is responsible for
the energy gap in the quasiparticle excitations.
Finally we find that
the presence of shadow peaks above the Fermi surface
can essentially
occur as a result of the singlet bonding on the rungs,
indicating that the on-site Coulomb repulsion
along chains or rungs 
may not be a prime cause
for the formation of shadow peaks.

One of us (S.H.S.S) acknowledges the financial supports
of Korean Ministry of Education (BSRI-97) 
and of the Center for Molecular Sciences at KAIST.

\newpage
\centerline{FIGURE CAPTIONS} 
\begin{itemize}
 \item[FIG. 1.]
  The quasiparticle energy dispersion obtained 
  by using the mean field approach in the weak coupling limit.
  The parameters are chosen
  from our Lanczos calculations on a $2\times6$ ladder with two doped holes as
  $t_\parallel=t_\perp=1$, $J=0.4$, $\chi=0.47$, $\Delta=-0.15$, $\mu=-0.099$.
  B denotes a bonding ($k_y=0$) band and
  A, an antibonding ($k_y=\pi$) band.
  The dashed parts of the dispersions
  denote shadow bands.
 \item[FIG. 2.]
  Exact diagonalization results of the hole spectral function
  on a $2\times6$ ladder with two doped holes
  for $t_\parallel=t_\perp=1$, $V=0$,
  and several values of $U$ and $J$.
  The dashed horizontal line denotes the Fermi energy,
  and the dotted arrows indicate the shadow peaks.
  The $\delta$ functions have been given a finite width of $\epsilon=0.1$.
 \item[FIG. 3.]
  Same as in Fig. 2 but for $J=0$.
 \item[FIG. 4.]
  Same as in Fig. 2 but for $U=0$.

\end{itemize}

\newpage

\newpage

\newpage

\newpage

\end{document}